\documentclass[%
 reprint,
superscriptaddress,
 amsmath,amssymb,
 aps,
 prl,
]{revtex4-2}

\usepackage{graphicx}
\usepackage{dcolumn}
\usepackage{bm}
\usepackage{svg}
\usepackage{microtype} 
\usepackage{paralist}
\usepackage[normalem]{ulem}
\usepackage[unicode=true, colorlinks=true, citecolor={blue!80!black}, urlcolor={blue!50!black}, linkcolor = {blue!80!black}]{hyperref} 
\usepackage{array}
\usepackage{siunitx}
\usepackage{wrapfig}
\usepackage{enumitem}
\usepackage{amsmath,amssymb}
\usepackage{array}
\usepackage{soul}
\usepackage{centernot}
\usepackage{bibunits}
\usepackage[titletoc, title]{appendix}
\usepackage{titletoc}
\usepackage{lineno}
\usepackage{lipsum}  
\usepackage{printlen}

\newcommand*{\balancecolsandclearpage}{%
  \cleardoublepage
  \twocolumngrid
}

\newcommand{\ket}[1]{\left| #1 \right>}
\newcommand{\bra}[1]{\left< #1 \right|}

\newcommand{\sx}[1]{\sigma_{#1,x}}
\newcommand{\sy}[1]{\sigma_{#1,y}}
\newcommand{\sz}[1]{\sigma_{#1,z}}

\begin{document}

\preprint{APS/123-QED}

\title{Kramers-protected hardware-efficient error correction with Andreev spin qubits}

\author{Haoran Lu}
\thanks{These two authors contributed equally}
\affiliation{School of Applied and Engineering Physics, Cornell University, Ithaca, NY, 14853, USA}

\author{Isidora Araya Day}
\thanks{These two authors contributed equally}
\affiliation{ QuTech, Delft University of Technology, Delft 2600 GA, The Netherlands}
\affiliation{Kavli Institute of Nanoscience, Delft University of Technology, 2600 GA Delft, The Netherlands}

\author{Anton R. Akhmerov}
\affiliation{Kavli Institute of Nanoscience, Delft University of Technology, 2600 GA Delft, The Netherlands}

\author{Bernard van Heck}
\affiliation{Dipartimento di Fisica, Sapienza Università di Roma, Piazzale Aldo Moro 2, 00185 Rome, Italy}

\author{Valla Fatemi}
\email{vf82@cornell.edu}
\affiliation{School of Applied and Engineering Physics, Cornell University, Ithaca, NY, 14853, USA}

\date{\today}

\begin{abstract} 
We propose an architecture for bit-flip error correction of Andreev spins that is protected by Kramers' degeneracy. 
Specifically, we show that a coupling network of linear inductors and Andreev spin qubits results in a static Hamiltonian composed of the stabilizers of a bit-flip code. 
The electrodynamics of the many-body spin states also respect these stabilizers, and we show how reflectometry off a single coupled resonator can thereby accomplish their projective measurement.
We further show how circuit-mediated spin couplings enable error correction operations and a complete set of single- and two-module logical quantum gates.
The concept, which we dub the Ising molecule qubit (or Isene), is experimentally feasible and provides a path for compact noise-biased qubits. 
\end{abstract}

\maketitle

\textit{Introduction --} Andreev spin states are a new qubit platform composed of microscopic spin degrees of freedom that couple to macroscopic supercurrents~\cite{chtchelkatchev_andreev_2003,padurariu_theoretical_2010,park_andreev_2017,tosi_spin-orbit_2019,hays_continuous_2020,hays_coherent_2021,bargerbos_spectroscopy_2023,pita-vidal_direct_2023,pita-vidal_strong_2024}. 
Such direct integration of spins with superconductivity raises the prospect for strong or ultrastrong coupling to other superconducting degrees of freedom~\cite{devoret_circuit-qed_2007}, including microwave-frequency resonators and qubits~\cite{bargerbos_spectroscopy_2023,pita-vidal_direct_2023,vakhtel_quantum_2023,vakhtel_tunneling_2024,pita-vidal_strong_2024}, and to other Andreev spins~\cite{pita-vidal_strong_2024}. 
One key consequence, as pointed out in a recent blueprint~\cite{pita-vidal_blueprint_2025}, is that Andreev spins can be the basis of a solid-state qubit platform endowed with fast-tunable, all-to-all connectivity. 
All-to-all connectivity can significantly reduce the overhead for both quantum error correction~\cite{chamberland_topological_2020,bravyi_quantum_1998,dennis_topological_2002,bravyi_high-threshold_2024} and analog quantum simulation applications~\cite{lechner_quantum_2015,baumer_efficient_2024}. 

Here, we show how Kramers' theorem~\cite{kramers_theorie_1930} can gainfully constrain the spin-spin couplings by guaranteeing the absence of couplings involving odd numbers of spins.
We propose a hardware-efficient implementation of bit-flip error detection that relies on a single readout resonator by leveraging the many-body electrodynamics of the spin array. 
Local spin control then allows error correction while the logical basis states remain degenerate.
Multi-frequency drives and fast flux control further provide a complete set of logical gates. 
Experimentally demonstrated hardware is capable of accomplishing a demonstration of our proposal, provided future improvements in spin dephasing times~\cite{hays_coherent_2021,pita-vidal_direct_2023}.
In the main text we describe the salient aspects of our concept, while appendices and published code provide technical details.

\begin{figure}[t]
\centering
\includegraphics[width = 0.85\columnwidth]{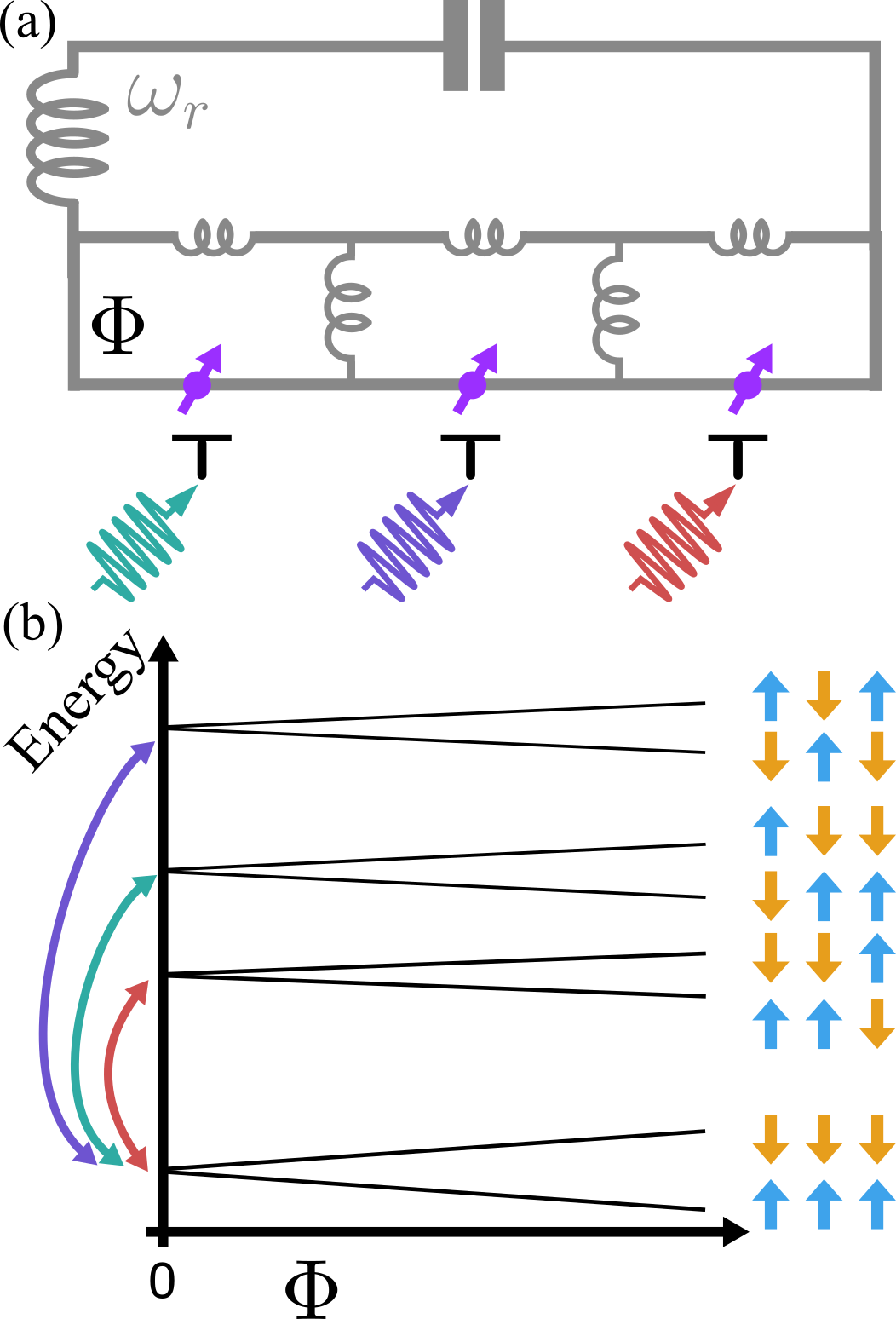}
\caption{ 
(a) Circuit for minimal bit-flip encoding with Andreev spins.
The field-effect gate lines on each spin (black) include DC gate voltages (not shown) and AC drive (colored). 
Control circuitry for flux $\Phi$ is not shown.
Andreev spins are modeled by Eq.~\eqref{eq:ASQ_EPR}, while linear inductors by a quadratic energy-phase relation $U_L = (\varphi_0^2/2L)  \varphi^2$, with $\varphi_0 = \hbar/2e$ the reduced flux quantum.
(b) Schematic level structure as a function of flux, with the Kramers' point at zero flux. 
Colored arrows correspond to EDSR-induced single-spin transitions between the logical manifold (lowest) and the error manifolds while they remain degenerate.
}
\label{fig:intro} 
\end{figure}

\textit{The Setup --}  The Andreev spin can be modeled as a Josephson weak link with a spin-dependent energy-phase relation~\cite{padurariu_theoretical_2010},
\begin{equation}
    U_\sigma = E_0 \cos(\varphi) + E_\sigma \sigma_z \sin(\varphi)~, \label{eq:ASQ_EPR}
\end{equation}
where $E_0$ describes the spin-independent contribution (and may be positive or negative), $E_\sigma>0$ describes the magnitude of the spin-dependent contribution, $\sigma_z$ is the Pauli $z$ matrix for the spin, whose orientation is defined locally, and $\varphi$ is the phase drop across the weak link. 

For this work, we focus on a minimal circuit hosting three Andreev spins in series, as shown in Fig.~\ref{fig:intro}(a). 
The spin-dependent supercurrent from each spin biases the other spins, which results in spin-spin interactions. 
Attachment of linear inductors to all nodes eliminates circuit offset charges, which may contribute to setpoint drift~\cite{vool_introduction_2017}.
The series configuration of spins is convenient for avoiding wiring crossovers and for controlling the range of interactions.
The depicted field-effect gate electrodes serve two roles: tuning up the spins and accomplishing fast ($\sim\SI{20}{\nano\second}$) and local spin control through electric dipole spin resonance (EDSR)~\cite{padurariu_theoretical_2010,metzger_circuit-qed_2021,bargerbos_spectroscopy_2023,pita-vidal_direct_2023,lu_andreev_2025}.
Finally, an attached resonator with average frequency $\omega_{r,0}$ results in one dynamical phase degree of freedom across the entire array, which we will show can be used for simultaneous readout of the three Andreev spins.
We implemented an algorithm (detailed in the Appendix) to determine the energy of each spin configuration and their influence on the resonator.
We note that the basic physics described here is also compatible with parallel arrays and with nonlinear inductive couplers~\cite{padurariu_theoretical_2010,pita-vidal_strong_2024,pita-vidal_blueprint_2025}.

\textit{The Effective Ising Hamiltonian --}  With this setup, we are now ready to inspect the Hamiltonian. 
For generic values of the circuit parameters, the eight distinct spin configurations all have distinct energies~\cite{padurariu_theoretical_2010,pita-vidal_blueprint_2025} and an effective Hamiltonian of the circuit contains all possible Z-type couplings between any number of spins.
If the magnetic fluxes in all loops are zero or $\pi$, a time-reversal symmetric configuration, Kramers' theorem guarantees that any coupling involving an odd number of spins vanishes~\cite{kramers_theorie_1930}. 
The following static Hamiltonian results: 
\begin{equation}
    H_\sigma = J_{12}\sz{1}\sz{2} + J_{23}\sz{2}\sz{3} + J_{13} \sz{1}\sz{3}~. \label{eq:ZZHam}
\end{equation}
Because of the similarity to the Ising model, coupled with the finite size consistent with a small molecule, we dub this an Ising molecule, or Isene. 
The specific values and signs of the spin-spin couplings $J_{ij}$ depend on the details of the circuit and weak links, and so they are designable.
When the inductors and ASQs are not identical, the $J_{ij}$ will be different: for our purposes, this is a desired feature as it eliminates accidental degeneracies. 
We note in particular that this Hamiltonian form scales to more than three Andreev spin qubits (with additional higher-order interactions involving even numbers of spins) and that the range of the interaction depends on the vertical inductances.  
We also expect double degeneracy of the full spectrum to be protected from charge noise-induced dephasing, mitigating a common concern about spin-orbit type qubits~\cite{san-jose_spin_2007}.
Finally, coupled-qubit modules based on bosonic degrees of freedom, e.g.~\cite{roy_programmable_2020}, do not have a generic way to remove the single- and three-spin interactions.

We recognize the Ising Hamiltonian~\eqref{eq:ZZHam} as being comprised of the stabilizers of a bit flip error correction code. 
The eigenstates are pairwise degenerate as guaranteed by Kramers' theorem, and the excited states are connected to the ground state by single spin flips as shown in Fig.~\ref{fig:intro}(b). 
We remark that the Hamiltonian~\eqref{eq:ZZHam} was also described for parallel arrays~\cite{padurariu_theoretical_2010,pita-vidal_blueprint_2025}, but the physics of operation at Kramers' point as the idling point was not considered.

\textit{Error Detection and Correction --}  We now go further to consider the linear response to the phase drop across the array. 
We focus here on the inductance of the circuit that terminates a resonator, Fig.~\ref{fig:intro}(a).
The Andreev spin configuration affects the resonator frequency, resulting in contributions to the Hamiltonian at Kramers' point given by 
\begin{align}\label{eq:readoutHam}
    H_{r}  =&  \hbar  \omega_{r,0}  \hat{a}^\dagger \hat{a} +\sum_{i\neq j} \hbar\chi_{ij}\sz{i}\sz{j}\,\hat{a}^\dagger \hat{a}\,.
\end{align}
where $\hat{a}$ is the lowering operator for the resonator, $\omega_{r,0}$ is the spin-unconditional resonator frequency, and $\chi_{ij}$ are spin-conditional frequency shifts. 

The total static Hamiltonian is then
$H=H_\sigma + H_r$. We note that  $H_r$ respects the same pairwise structure of spin-spin couplings as $H_\sigma$.
Therefore, a microwave reflectometry measurement commonly used on Andreev spins and superconducting qubits~\cite{hays_continuous_2020,metzger_circuit-qed_2021,fatemi_microwave_2022,blais_circuit_2021,kjaergaard_superconducting_2020} would accomplish projective measurement of the stabilizers, provided that the shifts $\chi_{ij}$ are different.
We note that, as before, Kramers' theorem guarantees that only coupling terms between two spins appear in $H_r$. 
The absence of single-spin terms avoids the need to fine-tune the matching of resonator shifts due to individual qubits in order to perform error detection, as necessary with e.g. superconducting qubit architectures~\cite{liu_comparing_2016,andersen_entanglement_2019,roy_programmable_2020,livingston_experimental_2022}.

This scheme also allows error correction without deviating from Kramers' point. 
The many-body spin configurations related by a \textit{single} spin flip are not time-reversed partners, so Kramers' theorem does not forbid those transitions.  
Indeed, a local EDSR drive to spin $j$ accomplishes 
\begin{equation}
    H_{d,j} \approx M_j \sy{j} \sum_{k\neq j} A_{kj} \sz{k}~, \label{eq:drive}
\end{equation} 
where $M$ encloses a drive amplitude, and $A_{kj}$ are matrix elements proportional to spin-induced phase shifts: 
$A_{kj}$ is nonzero because spin-spin interactions mediated by the inductors ensure that the equilibrium value of the phase difference across each individual junction is nonzero and depends on the spin configuration. 
This results in a nonzero matrix element to EDSR control (see Appendix for details).
Therefore, the stabilizer readout may be followed by a single EDSR $\pi$ pulse on the relevant spin to restore the system to the logical manifold.
We remark that our scheme can extend beyond three spins to $N$ spins for higher-order bit flip error correction.

We can now compare our concept with implementations of the bit-flip code that use conventional qubits~\cite{riste_detecting_2015, riste_real-time_2020}. 
There, error detection of a single bit-flip error in three physical data qubits requires four cNOT gates to two ancilla qubits, which must both then be read out.
In comparison, our protocol requires a single readout, no gates, and no ancilla qubits; hence, we consider our approach as hardware-efficient. 
In both cases, error correction requires a single $\pi$-pulse conditioned on the stabilizer measurement. 

\begin{figure}
\centering
\includegraphics[width = 0.99\columnwidth]{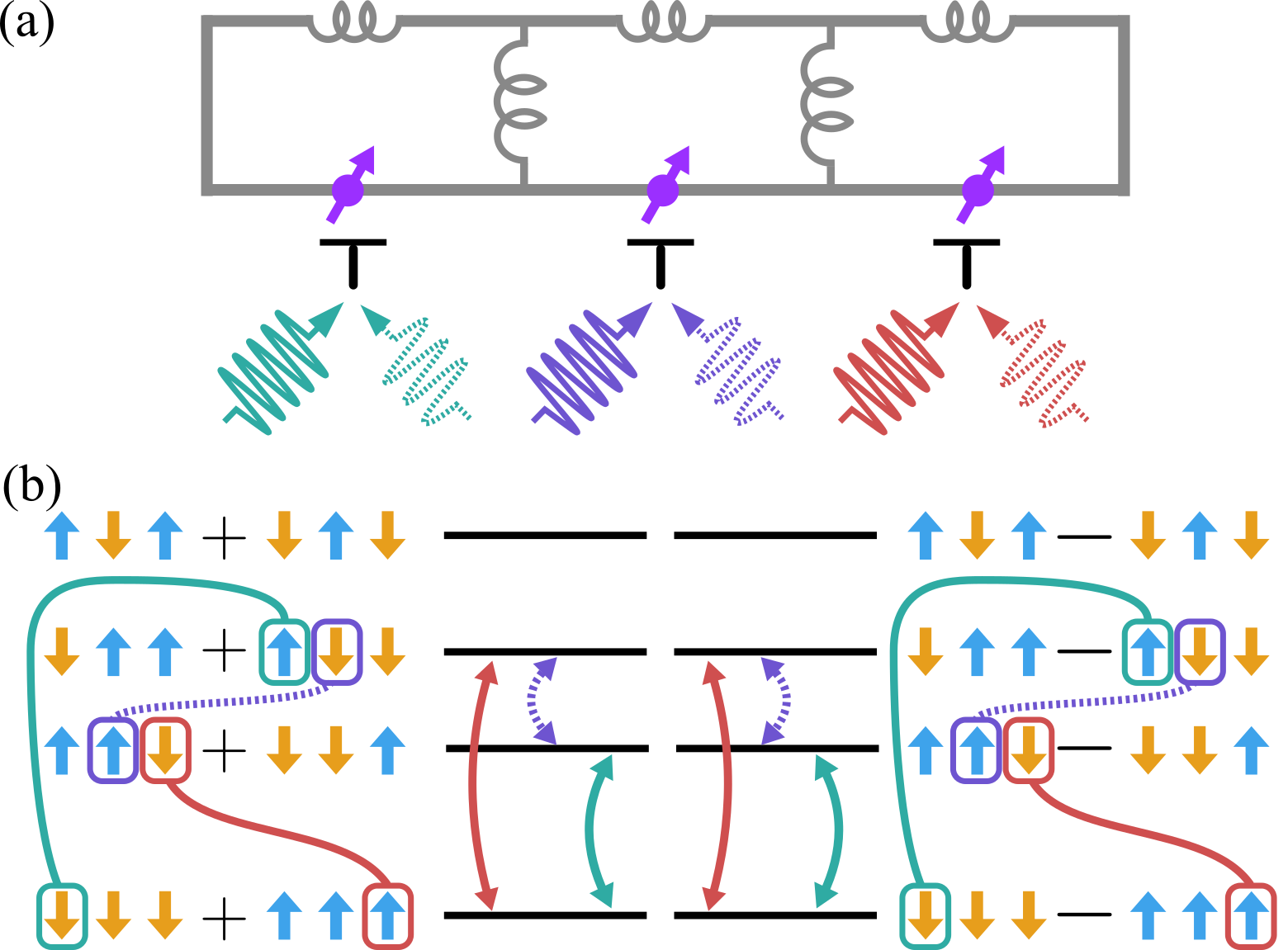}
\caption{ 
(a) Each spin can be driven locally via EDSR. The two tones depicted imply two possible frequencies for resonant transitions. The solid line is the higher frequency drive.
(b) Operations that invoke three or more states that complete a loop in state space can accomplish logical $X(\theta)$ gates. 
Three sequential $\pi$ pulses at the transitions shown would accomplish a logical $X(\pi)$ gate. 
See main text for other $\theta$.
We have chosen here a gauge where matrix elements for the symmetric eigenstates of $X$ are equal to the negative of the antisymmetric eigenstates. 
}
\label{fig:XXXrot} 
\end{figure}

\textit{Logical Quantum Gates --} We now describe the available logical quantum gates. 
A total Hamiltonian containing the static part~\eqref{eq:ZZHam} and drives of the form~\eqref{eq:drive} commutes with the operator $X = \sx{1} \sx{2} \sx{3}$. 
Therefore, the manifolds can be decomposed into decoupled symmetric ($+$) and antisymmetric ($-$) subspaces of $X$, as shown in Fig.~\ref{fig:XXXrot}(b). 
The static Hamiltonian,~\eqref{eq:ZZHam} and~\eqref{eq:readoutHam}, and the drive Hamiltonian~\eqref{eq:drive} are block-diagonal in this basis:
\begin{equation}
    H = 
    \begin{pmatrix}
    H_{\sigma}^+ + H_{r}^+ + \sum_j H_{d,j}^+& 0 \\
    0 & H_{\sigma}^+ + H_{r}^+ -  \sum_j H_{d,j}^+
    \end{pmatrix}
\end{equation}
The opposite sign for the driving terms in the two subspaces, $H_{d,j}^+ = -H_{d,j}^-$, will enable logical $R_X(\theta)$ gates.

While the use of one spin transition alone cannot achieve a logical gate,
we find involvement of three states sufficient to achieve a relative logical phase in the $X$ basis. 
Following the sequence of transitions indicated in Fig.~\ref{fig:XXXrot}(b), and assuming on-resonant rotations within the rotating wave approximation (RWA), three sequential $\pi$ pulses accomplishes $R_X(\pi)$.
To accomplish arbitrary $\theta$,  one can instead start with rotations of $\pi/2$ and $\pi$ for the first two pulses to accomplish a manifold-independent Hadamard operation between the first and third level. 
Then one can apply a drive between those same levels with the operator $\pm (\ket{0\pm}\bra{2\pm} + \ket{2\pm}\bra{0\pm})$, where $\pm$ refers to the symmetric and antisymmetric subspaces, and $\ket{0\pm}$ and $\ket{2\pm}$ refer to the first and third states. 
This drive is used to accomplish a relative phase $\theta$ based on the rotation angle.
Then one accomplishes the inverse of the first two pulses to return to the original manifold.  
Therefore, any logical rotation $R_X(\theta)$ is possible.

In real systems, operations that rely on  RWA may require pulses that are too long in duration. 
Using Krotov optimization~\cite{goerz2019}, we further found diabatic operations with time-dependent $M_j$ that accomplish $R_X(\theta)$ for $\theta = \{\pi/2, \pi/4, \pi/8$\}, with arbitrary precision (absent nonidealities like decoherence).
The details of this procedure are given in the Appendix.
Therefore, we expect that a continuous range of logical rotation angles with fast driving are possible, using pulses derived by optimal control methods.

Additionally, a continuous logical phase gate $R_Z(\theta)$ can be accomplished by local flux control. 
Consider the dispersion in flux shown in Figure~\ref{fig:intro}(b).
A local flux pulse $\Phi(t)$ detunes the energy difference $\Delta E(t)$ between the many-body spin configurations within (and between) the computational manifolds. 
By ensuring that the flux starts and ends at zero, $\Phi(0)=\Phi(T)=0$, the total accumulated logical phase difference is $\theta = \hbar^{-1} \int_0^T \left[ E_{\uparrow\uparrow\uparrow}(\Phi(t)) - E_{\downarrow\downarrow\downarrow}(\Phi(t)) \right] dt$.
We note that, in principle, this does come at the cost of breaking the manifold degeneracies afforded by Kramers' theorem while the gate is being applied.
Nonetheless, continuous logical $R_X(\theta)$ and $R_Z(\theta)$ are sufficient for universal quantum control of the logical manifold of the module. 

\begin{figure}[b]
\centering
\includegraphics[width = 0.99\columnwidth]{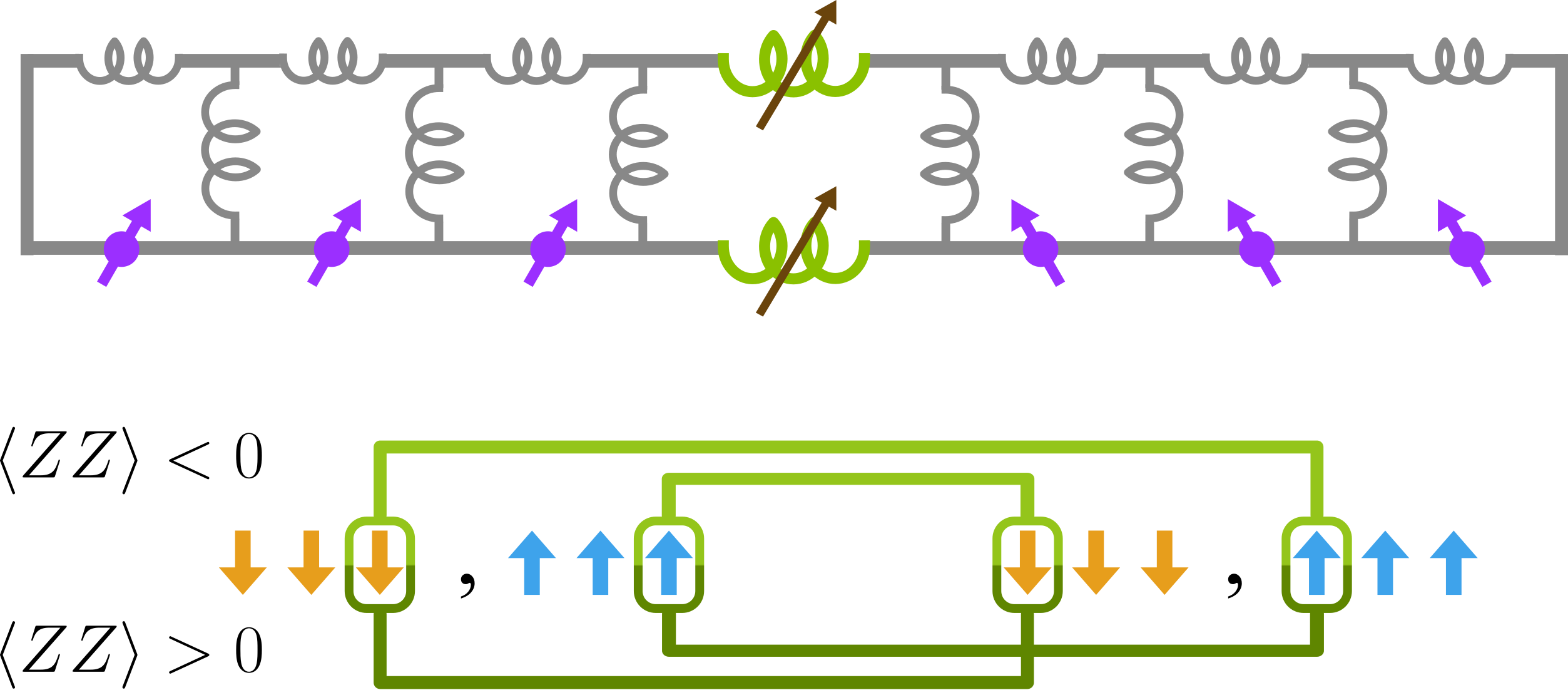}
\caption{ 
Tunable inductors (inductor symbols with arrows) can used to turn on/off a spin-spin interaction between the neighboring spins of two modules.
The spin-spin interaction of the form $\sz{3,a}\sz{1,b}$ on the last spin spin of module $a$ and the first spin of module $b$ (indicated by the green boxes) translates to a logical $ZZ$ interaction. 
}
\label{fig:twomodule} 
\end{figure}

Finally, a two-module logical $R_{ZZ}(\theta)$ gate can be accomplished by turning on/off an inductive coupling between the modules, schematically shown in Fig.~\ref{fig:twomodule}. 
We denote the two modules with $a$ and $b$. 
When the inductive coupling is activated, the dominant spin-spin interaction will be nearest-neighbor, or $J_{3a,1b}\sz{3a}\sz{1b}$. 
The intermodule spin interaction should be set to be weaker than the intramodule spin interactions, $J_{3a,1b} < J_{ij,a},J_{ij,b}$, to ensure the logical states remain well separated from the error states. 
As a result, this interaction accomplishes a logical $R_{ZZ}(\theta)$ gate with $\theta$ determined by the area $\theta = \hbar^{-1} \int_0^T 2 J_{3a,1b}  dt$.
The tunable inductance could be accomplished either with flux-modulated SQUIDs or gate-modulated superconductor-semiconductor Josephson junctions~\cite{yan_tunable_2018,oconnell_yuan_epitaxial_2021}.

\textit{Feasibility and Outlook --} The experimental feasibility of our concept requires balancing the energy scales of the Andreev spins in Eq.~\eqref{eq:ASQ_EPR} with the characteristic energy scales of the inductors, $\varphi_0^2 / 2 L$. 
Typical measured values for $E_0$ and $E_\sigma$ range from 0.1 to 1 \si{\giga\hertz}~\cite{tosi_spin-orbit_2019,hays_continuous_2020,hays_coherent_2021,metzger_circuit-qed_2021,bargerbos_spectroscopy_2023,pita-vidal_direct_2023,pita-vidal_strong_2024,lu_andreev_2025}.
We find that by using inductors of the scale  1 to 10 $\si{\nano\henry}$, which are commonly implemented~\cite{hays_continuous_2020,samkharadze_high-kinetic-inductance_2016,pita-vidal_gate-tunable_2020,lu_andreev_2025}, we can achieve couplings $J_{ij}/2\pi$ at the $\si{\mega\hertz}$ scale.
This is sufficient for gates as fast as $\sim\SI{100}{\nano\second}$ with baseband control.
We note that the required inductance is inversely proportional to $E_\sigma$.
Modeling the coupling to a typical distributed-element resonator from previous Andreev spin experiments~\cite{hays_continuous_2020,lu_andreev_2025}, the resonator frequency shifts $\chi_{ij}/2\pi$ are predicted to be 0.1 to 1~\si{\mega\hertz}, which is a standard range for single-shot dispersive readout in a similar duration.
Finally, we estimate the EDSR-activating parameter $A_{kj} \gtrsim 0.01$, sufficient for gate driving. 
We note that by entering the non-perturbative circuit coupling regime in which the coupling inductances are comparable to the weak link inductance (such as proposed and recently accomplished in parallel arrays with a Josephson junction coupler~\cite{padurariu_theoretical_2010,pita-vidal_strong_2024,pita-vidal_blueprint_2025}), the spin-spin interaction strengths $J_{ij}/2\pi$ can reach the \SI{100}{\mega\hertz} scale, two orders of magnitude larger. 
In this regime, the circuit may become highly nonlinear and strongly hybridized with the spin states, and so an additional mode may need to be added for convenient readout~\cite{pita-vidal_strong_2024}.
On the other hand, it carries the benefit that the $J_{ij}$ approach the thermal energy scale, which could suppress single spin errors. 

The decoherence rates of the Andreev spins pose a challenge.
The existing experimental implementations for Andreev spins rely on the semiconductor InAs, in which confined electronic spins are consistently observed to dephase at the \SI{10}{\nano\second} scale~\cite{nadj-perge_spinorbit_2010,hays_coherent_2021,pita-vidal_direct_2023}. 
One main suspect for this dephasing is the nuclear spin bath of InAs~\cite{hoffman_decoherence_2024}, which should result in a similar scale of decoherence at Kramers' point. 
Therefore, our proposal can serve as a target for next-generation platforms for Andreev spins, such as in isotopically purifiable germanium~\cite{scappucci_germanium_2021,lakic_proximitized_2024} and carbon~\cite{park_steady_2022}. 
Charge noise is also a concern~\cite{hoffman_decoherence_2024}, for which our proposal should provide first-order protection at the idling point.

We also speculate on how one might extend to full quantum error correction. 
The module proposed here allows for a computationally complete set of single- and two-module logical quantum gates while actively correcting one type of error. 
The $R_Z(\theta)$ and $R_{ZZ}(\theta)$ gates commute with the dominant error channel (dephasing), while the $R_X(\theta)$ gate does not. 
As a result, the module has the potential to be a noise-biased qubit embedded in a larger encoding. 
This is attractive because noise-biased qubits benefit from improved scalability in quantum error correction~\cite{aliferis_fault-tolerant_2008,tuckett_ultrahigh_2018,guillaud_repetition_2019,bonilla_ataides_xzzx_2021,chamberland_building_2022,hann_hybrid_2024}.
Unlike noise-biased qubits based on bosonic cat codes in superconducting oscillators~\cite{vlastakis_Deterministically_2013,grimm_stabilization_2020,putterman_hardware-efficient_2025}, Andreev spin qubits are microscopic in size. 
However, $X$ and controlled-$X$ gates that commute with the dominant error channel are considered necessary to unlock increased thresholds of noise-biased encodings.
A search for such gates will be left for later work. 
    


Finally, we remark on possible connections with an existing three-spin module, the exchange-only qubit~\cite{russ_three-electron_2017}. 
There, the three spins are coupled through the exchange interaction; this interaction is of the isotropic Heisenberg form when spin-orbit coupling is not considered, which has been found to be reasonable for spins hosted in silicon.
This interaction generically leaves behind a four-fold degeneracy, so it is unlikely to be useful for single-spin error detection and correction. 
In contrast, the supercurrent-based interaction of our proposal constrains the spin-spin couplings to be along one axis only.
However, for spins in germanium, the stronger spin-orbit interaction has been observed to induce anisotropic exchange interactions~\cite{katsaros_zero_2020,hetenyi_anomalous_2022}, so an analogue of our proposal may be possible without superconductors if an extreme anisotropy can be engineered. 
\\

All data generated and code used in this work are available at: 10.5281/zenodo.14277238.

\begin{acknowledgments}
V. F. acknowledges helpful discussions with Baptiste Royer, Nicholas Frattini, and Chao-Ming Jian, as well as an early conversation with Christian Andersen on parity readout. 
We thank Christian Andersen, Pavel Kurilovich, Baptiste Royer, and Shyam Shankar for input on the manuscript and the concept. 
We thank Stefano Bosco for education and references on the case of normally-confined spin qubits.

Research was sponsored by the Army Research Office and was accomplished under Grant Number W911NF-22-1-0053. The views and conclusions contained in this document are those of the authors and should not be interpreted as representing the official policies, either expressed or implied, of the Army Research Office or the U.S. Government. The U.S. Government is authorized to reproduce and distribute reprints for Government purposes notwithstanding any copyright notation herein.
Research by I.A.D. and A.A. was supported by the Netherlands Organization for Scientific Research (NWO/OCW) as part of the Frontiers of Nanoscience program, a NWO VIDI grant 016.Vidi.189.180.
We also thank the Global Quantum Leap sponsored by the National Science Foundation AccelNet program under award number OISE-2020174 for funding an exchange visit by I. A. D. 

\end{acknowledgments}

V. F. conceived the project and coordinated the research.
H. L. and B. vH. accomplished initial calculations, and B. vH. wrote down the general analytical approach.
I. A. D. and A. A. developed the numerical methods for solving the inductive circuit and performed the Krotov optimization, as well as identified the $X$ symmetry. 
H. L. developed code for calculating the resonator frequency shifts.
H. L. and I. A. D. calculated results for the circuits shown in the main text.
V. F. wrote the manuscript with input from all authors.
The authors declare no competing interests.



\appendix

\newpage

\section{Appendix}

\textbf{Solving the circuit}\\
The readout technique of the Andreev spins relies on the spin-dependent inductance of the circuit~\cite{park_adiabatic_2020,kurilovich_microwave_2021,fatemi_microwave_2022}:
\begin{equation}
    E_L = \varphi_0^{2} \frac{\partial^2 E_g}{\partial \Phi_{\textrm{in}}^2} \Bigg     \rvert_{\Phi^\ast_{\textrm{in}}},
    \label{eq:inductance_derivative}
\end{equation}
where $E_g$ is the ground state energy of the circuit,  $E_L$ is the inductive energy scale proportional to the inverse inductance, $\Phi_{\textrm{in}}$ is the incoming flux through the circuit, and $\Phi^\ast_{\textrm{in}}$ indicates the value of $\Phi_{\textrm{in}}$ at the energetic minimum.
We note that we neglect virtual spin transitions~\cite{kurilovich_microwave_2021} which can also contribute but are expected to be weak~\cite{hays_coherent_2021}. 
In this section, we describe how to compute the energies of the $8$ spin configurations and their respective inductances.

\begin{figure}
\centering
\includegraphics[width = 0.8\columnwidth]{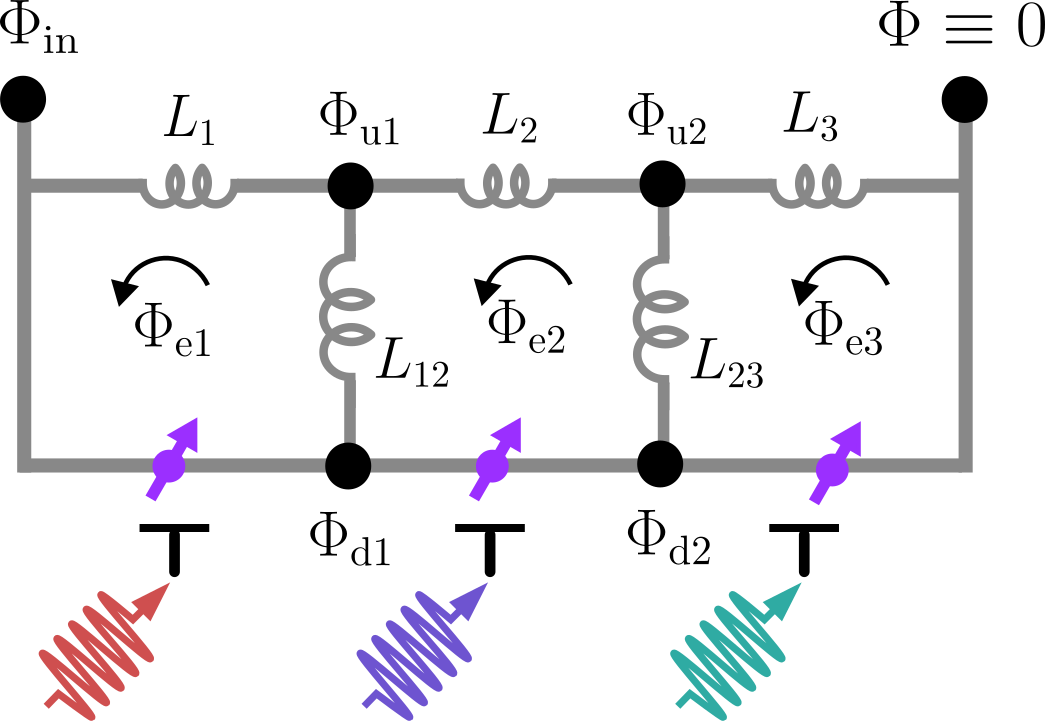}
\caption{The circuit diagram labeling the individual inductors, external fluxes, and circuit fluxes at each node.
The gates at the bottom  (black T shapes) can be driven electrically at nonzero frequency (wavy arrows). 
}
\label{fig:SI_circuit} 
\end{figure}

To solve the circuit, we first write its energy:
\begin{equation}
\mathcal{V} = 
\sum_{i=1}^3 \left( U_i\left( \frac{\Phi_{B_i}}{\varphi_0} \right) + \frac{\Phi_{T_i}^2}{2L_i} \right) + \frac{\Phi_{M_1}^2}{2L_{12}} + \frac{\Phi_{M_2}^2}{2L_{23}},
\label{eq:circuit_potential}
\end{equation}
where $i$ is the index of each Andreev spin qubit, $U_i(\phi) = E_{J,i} \cos(\phi) + E_{\sigma, i} \sigma_{z, i} \sin(\phi)$ is the Josephson potential of the weak link, and $L$ are the inductances.
The branch fluxes $\Phi_{T_i}$, $\Phi_{B_i}$, and $\Phi_{M_i}$ label the top, bottom, and middle branches of the circuit, respectively.
In terms of the node fluxes, the branch fluxes are:
\begin{subequations}
\begin{align}
\Phi_{T_1} &= \Phi_{\textrm{in}} - \Phi_{u_1} -\Phi_{e_1}\,, &\Phi_{B_1} &= \Phi_{\textrm{in}} - \Phi_{d_1}\,,\\
\Phi_{T_2} &= \Phi_{u_1} - \Phi_{u_2} - \Phi_{e_2}\,, &\Phi_{B_2} &= \Phi_{d_1} - \Phi_{d_2}\,,\\
\Phi_{T_3} &= \Phi_{u_2} - \Phi_{e_3}\,, &\Phi_{B_3} &= \Phi_{d_2}\,,\\
\Phi_{M_1} &= \Phi_{u_1} - \Phi_{d_1}\,, &\Phi_{M_2} &= \Phi_{u_2} - \Phi_{d_2}\,,
\end{align}
\end{subequations}
where $\Phi_{\textrm{in}}$ is the input flux to the circuit, $\Phi_{u_i}$ and $\Phi_{d_i}$ are the node fluxes of the vertical branch nodes, and $\Phi_{e_i}$ are the external fluxes threaded through each loop (see Fig.~\ref{fig:SI_circuit}).
We input these node fluxes in Eq.~\eqref{eq:circuit_potential} and rescale them into phase variables, $\phi_i = \Phi_i/\varphi_0$, to get
%
\begin{equation}
\begin{split}
\mathcal{V} =
&-\sqrt{E_{0,1}^2+E_{\sigma1}^2}\cos\left(\phi_\textrm{in}-\phi_{d_1}-\gamma_1\sigma_{z}^{(1)}\right) \\
&- \sqrt{E_{0,2}^2+E_{\sigma2}^2}\cos\left(\phi_{d_1} - \phi_{d_2}-\gamma_2\sigma_{z}^{(2)}\right) \\
&- \sqrt{E_{0,3}^2+E_{\sigma3}^2}\cos\left(\phi_{d_2} -\gamma_3\sigma_{z}^{(3)}\right) \\
&+ \tfrac{1}{2} E_{L_1} (\phi_{\textrm{in}} -  \phi_{u_1} -\phi_{e_1})^2 \\
&+ \tfrac{1}{2} E_{L_2} (\phi_{u_1} - \phi_{u_2} - \phi_{e_2})^2 \\
&+ \tfrac{1}{2} E_{L_3} (\phi_{u_2} - \phi_{e_3})^2 + \tfrac{1}{2} E_{L_{12}} (\phi_{u_1}-\phi_{d_1})^2 \\
&+ \tfrac{1}{2} E_{L_{23}} (\phi_{u_2}-\phi_{d_2})^2~.
\end{split}
\end{equation}

To compute the inductance of the circuit using Eq.~\eqref{eq:inductance_derivative}, we need to find the energies of the eight spin configurations.
Because the energy of the circuit does not depend on dynamical variables, we find the ground state by minimizing the potential energy for each spin configuration:
\begin{equation}
\frac{\partial \mathcal{V}}{\partial \phi_{u_1}} = \frac{\partial \mathcal{V}}{\partial \phi_{u_2}} = \frac{\partial \mathcal{V}}{\partial \phi_{d_1}} = \frac{\partial \mathcal{V}}{\partial \phi_{d_2}} = 0.
\label{eq:euler_lagrange}
\end{equation}
This is justified if parasitic capacitances are all small enough~\cite{rymarz_consistent_2023}. 
This gives a set of four nonlinear coupled equations that determine the values of the internal nodes phase variables $\phi_{u_1}$, $\phi_{u_2}$, $\phi_{d_1}$, $\phi_{d_2}$ as a function of the external variables $\phi_{\textrm{in}}$, $\phi_{e_1}$, $\phi_{e_2}$, $\phi_{e_3}$.
We used analytical differentiation to compute exact solutions to Eq.~\eqref{eq:euler_lagrange}.
First, we compute the gradient of the potential energy with respect to the phase variables of the internal nodes and the incoming phase using the automatic differentiation library JAX~\cite{jax2018github}.
This gives the system of equations in Eq.~\eqref{eq:euler_lagrange} with an additional equation for the incoming phase.
Second, we solve the five nonlinear coupled equations using a root-finding algorithm~\cite{scipy_2020}, and find $\phi^\ast_{\textrm{in}}$, $\phi^\ast_{u_1}$, $\phi^\ast_{u_2}$, $\phi^\ast_{d_1}$, and $\phi^\ast_{d_2}$ that minimize $\mathcal{V}$.
Finally, we compute the inductive energy $E_L$ as the Schur complement of the Hessian of the potential energy $\mathcal{V},(\phi^\ast_{\textrm{in}}, \phi^\ast_{u_1}, \phi^\ast_{u_2}, \phi^\ast_{d_1}, \phi^\ast_{d_2})$.
This procedure successfully computes the spin-dependent energy and inductance of the circuit without numerical derivatives or numerical minimization.
The procedure is also straightforwardly extensible to additional spins and inductors. 

To obtain the dispersive shift, we simulate the Andreev circuit as a termination of a transmission line resonator.  
The resonance frequency is given by the smallest solution for $\omega_r$ in
\begin{equation}
 \cot\left(\frac{\omega_\mathrm{r} l}{v_\mathrm{eff}}\right)  = -\frac{i Z_\mathrm{L}}{Z_\mathrm{c}} ~,
\end{equation}
where $Z_\mathrm{c}$ is the characteristic impedance, $l$ is the length of the transmission line segment, $v_\mathrm{eff}$ is the light velocity in the transmission line, and $Z_\mathrm{L} = 2 i \omega_\mathrm{r} \varphi_0^{2} / E_L $ the impedance of the Andreev circuit obtained from Eq.~\eqref{eq:inductance_derivative}.  

\begin{figure}
\centering
\includegraphics[width = \columnwidth]{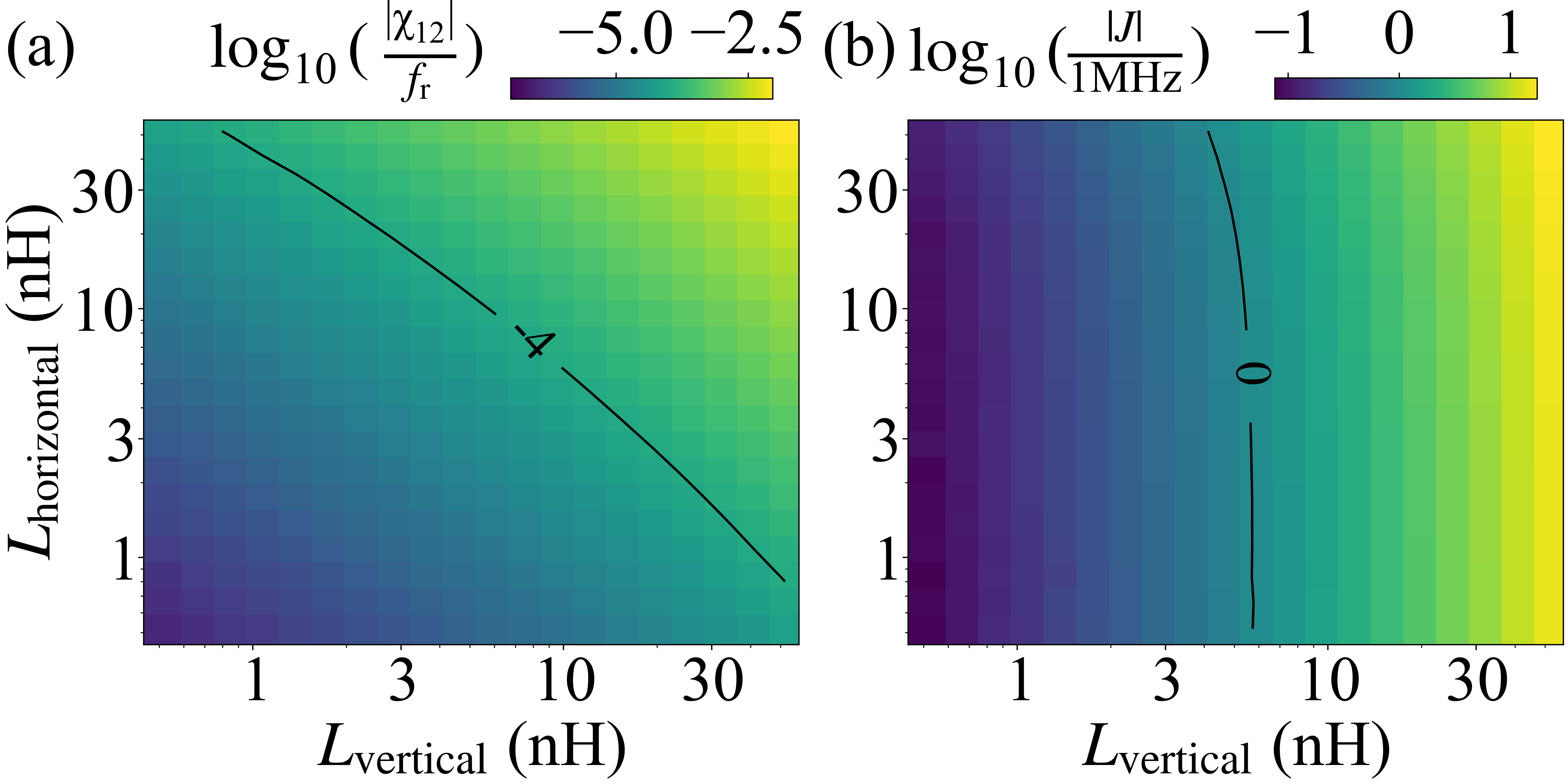}
\caption{(a) Dispersive shift $\chi_{12}$ and (b) inter-spin coupling $J_{12}$ at Kramers' point. Parameter used in the simulation: $E_{\sigma 1}/h = \SI{0.4}{\giga\hertz}$, $E_{\sigma 2}/h = \SI{0.3}{\giga\hertz}$, $E_{\sigma 3}/h = \SI{0.2}{\giga\hertz}$; $E_{0,1}/h = E_{0,2}/h = E_{0,3}/h = \SI{0.4}{\giga\hertz}$. $L_\mathrm{vertical} = L_\mathrm{12} = L_\mathrm{23}$ and $L_\mathrm{vertical} = L_\mathrm{1} = L_\mathrm{2} = L_\mathrm{3}$.
For this configuration $J_{23}$ and $\chi_{23}$ are of similar scale, while $J_{13}$ and $\chi_{13}$ are much smaller. 
}
\label{fig:SI_chi_J_2Dscan} 
\end{figure}

We obtain the dispersive shift contribution $\chi_{ij}$ and the inter-spin coupling energy $J_{ij}$ from the resonator frequency and ground state energies. 
We first assume a general form:

\begin{align}
   \begin{split}
    H =&  J_{0} + \sum_{h=1} J_{h}\sz{h} + \sum_{h \neq j}  J_{hj}\sz{h}\sz{j}\label{eq:H_Js}\\
    &\quad + J_{123}\sz{1}\sz{2}\sz{3},\\
   \end{split}\\
   \begin{split}
    H_{r}  =&  \hbar \hat{a}^\dagger \hat{a}  \left(   \omega_{r,0}+ 
        \sum_{h=1} \chi_{h}\sz{h} + \sum_{h\neq j} \chi_{hj}\sz{h}\sz{j}\right.\label{eq:H_chis}\\
    &\quad \Biggl.+ \chi_{123}\sz{1}\sz{2}\sz{3} \Biggr)~.\\
   \end{split}
\end{align}
With simulated data from all eight spin configurations, for $Z_c = \SI{50}{\ohm}$, $v_\mathrm{eff}= 0.39 c$ ($c$ is the free space speed of light), and varying $l \in \left[0.1,3.3\right]\si{\milli\meter}$ to fix $\omega_{r,0} = \SI{9}{\giga\hertz}$.
We then extract each coefficient in Eqs.~\eqref{eq:H_Js} and~\eqref{eq:H_chis}. 
We found that the odd-order terms indeed vanish at Kramers' point, bringing the Hamiltonian to the forms of Eq.~\eqref{eq:ZZHam} and Eq.~\eqref{eq:readoutHam}.
As an example calculation, Fig.~\ref{fig:SI_chi_J_2Dscan} displays $J_{12}$ and  $\chi_{12}$ as a function of the vertical and horizontal inductances (the inductors of the same orientation are uniform). 
It shows that the magnitude of the dispersive shift has strong dependence on both vertical and horizontal inductors, while the inter-spin coupling energy depends more on the vertical inductor.
\newline

\textbf{EDSR drive}\\
As pointed out by Padurariu and Nazarov, phase-biased Andreev spins can be driven by a finite-frequency drive on the gate~\cite{padurariu_theoretical_2010}. 
This is because the spin-orbit polarization pseudovector can depend on the gate voltage. 
Taking the case a single Andreev spin, they described the following drive Hamiltonian: 
\begin{equation}
     H(t) = \left[\vec{\epsilon}_{SO} +  \delta\vec{\epsilon}_{SO}\cdot \vec{\sigma} \cos(\omega t) \right] \sin(\hat{\varphi})~, \label{eq:EDSRfull}
\end{equation}
where $\vec{\epsilon}_{SO}$ is the unperturbed spin-orbit vector, $\delta\vec{\epsilon}_{SO}$ is the modulation of that vector due to the gate with the condition $|\delta\vec{\epsilon}_{SO}| \ll |\vec{\epsilon}_{SO}|$, and $\vec{\sigma}$ is the unperturbed Pauli matrix vector. 
Here we take a simplifying assumption that $\vec{\epsilon}_{SO} \propto \hat{z}$ and $\delta\vec{\epsilon}_{SO} \propto \hat{y}$, providing
\begin{equation}
     H(t) = \left[E_\sigma \sigma_z +  M \sigma_y \cos(\omega t) \right] \sin(\hat{\varphi})~. 
\end{equation}
For the single spin, shunted by an inductance, Kramers' point is at $\langle \hat{\varphi} \rangle = 0$, so that the EDSR drive cannot induce the transition. 
This method for manipulation of ASQs has been experimentally demonstrated~\cite{pita-vidal_direct_2023,lu_andreev_2025}.

When an additional spin is in the circuit, it induces a spin-dependent phase bias to the other spins. 
Taking the three-spin circuit of Figure~\ref{fig:intro}, when the weak link energies are much smaller than the inductive energies, we can approximate the average phase across spin $j$ as
\begin{equation}
    \langle \hat{\varphi}_j \rangle \approx \sum_k A_{jk}\sz{k}~,
\end{equation}
where the dimensionless weights $A_{jk}$ depend on the details of the circuit. 
Assuming weak spin-spin interactions giving $\langle \hat{\varphi}_j \rangle \ll 1$ (and consistent with simulations), we can Taylor expand the interaction \eqref{eq:EDSRfull} and apply the rotating wave approximation for $\hbar \omega = 2S $ to get Eq.~\eqref{eq:drive}, $H_{d,j} =  M \sy{j} \sum_k A_{jk} \sz{k}$.

Finally, we remark that the single-spin transition frequency is conditional on the state of the other spins. 
For the three-spin circuit, each spin is two spin-transition frequencies: one when the other two spins are aligned, and one where the other two spins are anti-aligned. 
To obtain an unconditional single-spin rotation, the drive must contain two frequency components resonant to the two transitions (with more frequencies for larger arrays). 
Multi-frequency control has been accomplished in solid state qubit contexts in several other situations~\cite{neeley_emulation_2009,champion_multi-frequency_2024,yu_creation_2024,roy_synthetic_2024}.\\

\textbf{Krotov optimization for $R_X(\theta)$}

\begin{figure}
\centering
\includegraphics[width = \columnwidth]{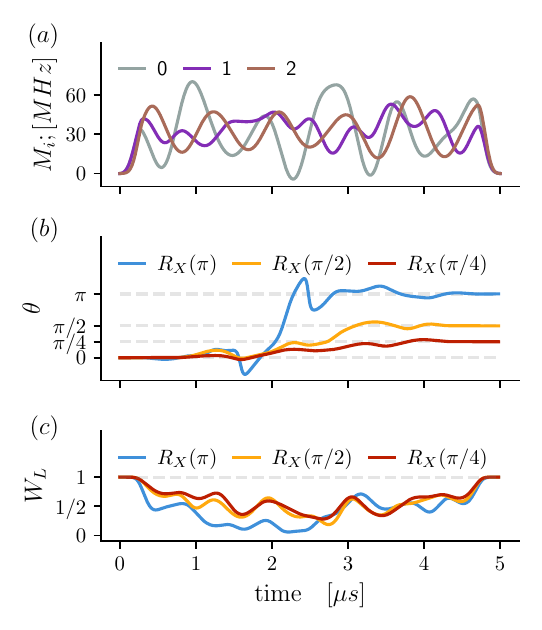}
\caption{Gate pulse optimization for $R_X(\theta)$ in the logical manifold.
(a) Optimized pulse amplitudes of the EDSR drives that achieve a $R_X(\pi / 2)$.
(b) Relative angle $\theta$ accumulated by the $\lvert + \rangle$ and $\lvert - \rangle$ states throughout the application of $\pi$, $\pi/2$, and $ \pi/4$.
(c) Weight of the time-evolved $\lvert 0 \rangle$ state on the logical manifold throughout the application of $\pi$, $\pi/2$, and $\pi/4$ rotations.
}
\label{fig:logical_sigma_x_gate} 
\end{figure}

We demonstrate that the EDSR drives are in principle enough to apply logical $R_x(\theta)$ rotations by performing numerical time-dependent simulations for three rotation angles $\theta = \pi$, $\theta=\pi/2$, and $\theta=\pi/4$.
To do this, we use the Kramers-protected Hamiltonian in Eq.~\eqref{eq:ZZHam} with the EDSR drive in Eq.~\eqref{eq:drive} with time-dependent amplitudes $M_j$.
We use the Python package Krotov~\cite{goerz2019} to find optimized pulse shapes of $M_1(t)$, $M_2(t)$, and $M_3(t)$ for the different $\theta$ objectives.
For the optimizations, we define the $R_X(\theta)$ gates over the logical manifold, and we set a maximum pulse duration of $\SI{5}{\micro\second}$, see Fig.~\ref{fig:logical_sigma_x_gate}(a) for the optimized pulses for $R_X(\pi/2)$, and the code in the repository for details on the other parameters.

As a result, a  time-evolved state $\lvert \Psi(t) \rangle$ is described by:
\begin{equation}
    \lvert \Psi(t) \rangle = \alpha_+(t) e^{i\phi_+(t)}\lvert + \rangle +  \alpha_-(t) e^{i\phi_-(t)} \lvert - \rangle + \sum_i \beta_i(t) \lvert \mathcal{E}_i \rangle,
\end{equation}
where $\alpha_+$ and $\alpha_-$ are the positive weights of $\Psi(t)$ on the $\lvert + \rangle$ and $\lvert - \rangle$ logical states.
The amplitudes $\beta_i$ are the weights on the rest of the Hilbert space states $\lvert \mathcal{E}_i \rangle$, the non-computational states.
Figure~\ref{fig:logical_sigma_x_gate}(b) shows the relative angle accumulated by the time evolution of the $\lvert + \rangle$ and $\lvert - \rangle$ states, $\theta(t) = \phi_+(t) - \phi_-(t)$, for the three example rotations.
Figure~\ref{fig:logical_sigma_x_gate}(c) shows the total weight of a time-evolved $\lvert 0 \rangle$ logical state on the logical manifold, $W(t) = \alpha_+^2(t) + \alpha_-^2(t)$.
All simulations use $500$ iterations of the Krotov algorithm.

\balancecolsandclearpage

\bibliography{vf-references,hl-references}
\end{document}